\documentclass[traditabstract]{aa}
\usepackage{txfonts}
\usepackage{graphicx}
\usepackage{color}
\usepackage{placeins}
\usepackage{natbib}

\newcommand\phs{\phantom{$-$}}
\newcommand{\rb}[1]{\raisebox{1.5ex}[-1.5ex]{#1}}

\begin{document}

\title{An outer shade of Pal: \\ 
Abundance analysis of the outer halo globular cluster Palomar 13}

\author{
Andreas Koch\inst{1},  
\and 
Patrick C\^ot\'e\inst{2}
}
\authorrunning{A. Koch \& P. C\^ot\'e}
\titlerunning{Abundance analysis of Palomar 13}
\offprints{A. Koch;  \email{andreas.koch@uni-heidelberg.de}}
\institute{
Zentrum f\"ur Astronomie der Universit\"at Heidelberg, Astronomisches Rechen-Institut, M\"onchhofstr. 12, 69120 Heidelberg, Germany 
\and
National Research Council of Canada, Herzberg Astronomy and Astrophysics Research Centre, 5071 West Saanich Road, Victoria, BC V9E 2E7, Canada
}
\date{}
\abstract{
At a Galactocentric distance of 27 kpc, Palomar~13 is an old globular cluster (GC) belonging to the outer halo.
We present a chemical abundance analysis of this remote system from high-resolution spectra obtained 
with the Keck/HIRES spectrograph. Owing to the low signal-to-noise ratio of the data, 
our analysis is based on a coaddition of the spectra of 18 member stars. We are able to 
determine integrated abundance ratios for 16 species of 14 elements, of 
$\alpha$-elements (Mg, Si, Ca, and Ti), Fe-peak (Sc, Mn, Cr, Ni, Cu, and Zn), and neutron-capture elements (Y and Ba). 
While the mean Na abundance is found to be slightly enhanced and halo-like, our method does not allow us  
to probe an abundance spread that would be expected in this light element if multiple populations are present in Pal~13. 
We find a metal-poor mean metallicity of $-1.91\pm0.05$ (statistical) $\pm$ 0.22 (systematic), confirming that Pal~13 is a typical metal-poor representative of the outer halo.
While there are some differences between individual $\alpha$-elements, such as halo-like Mg and Si versus the mildly lower Ca and Ti abundances, 
the mean [$\alpha$/Fe] of 0.34$\pm$0.06 is consistent with the marginally lower $\alpha$ component of the halo field and GC stars at similar metallicity. 
We discuss our results in the context of other objects in the outer halo and consider which of these objects were likely accreted. We also discuss the properties of their progenitors. 
While chemically, Pal~13 is similar to Gaia-Enceladus and some of its GCs, this is not supported by its kinematic properties within the Milky Way 
system. Moreover, its chemodynamical similarity with NGC 5466, a purported progeny of the Sequoia accretion event, might indicate a
common origin in this progenitor.
However, the ambiguities in the full abundance space of this comparison emphasize the difficulties in unequivocally labeling 
a single GC as an accreted object, let alone assigning it to a single progenitor.
}
\keywords{ Techniques: spectroscopic --- Stars: abundances -- Galaxy: abundances -- Galaxy: evolution -- Galaxy: halo -- globular clusters: individual: Palomar~13}
\maketitle
%
%
%%%%%%%%%%%%%%%%%%%%%%%%%%%%%%%%%%%%%%%%%%%%%%%%%%%%%%%%%%%%%%%%%%%%%%%%%%%%%
%
%
\section{Introduction}
The Galactic halo is a variegated place, populated by a mix of stellar populations and subsystems that likely have different origins. 
Our current understanding of hierarchical galaxy formation has been highly successful in 
identifying components that formed {\em \textup{in situ}} as well as those that formed in external environments, such as stars accreted from dwarf spheroidal (dSph) galaxies \citep{Cooper2013,PillePich2015} or
disrupted globular clusters (GCs), which have been shown to have contributed at the 
 $\sim$10\% level to the build-up of the stellar halo of the Milky Way (MW; \citealt{Martell2010}; \citealt{Koch2019CN}). 
In turn, \citet{Kruijssen2019} and \citet{Massari2019} estimated that $\sim$35--40\% (i.e., $\sim$55--65) 
of the present-day MW GCs are of extragalactic origin
and that the main contribution to the halo up to $z\sim2$ came from three major accretion events: Sagittarius \citep{Ibata1994,Law2010}, Gaia-Enceladus \citep{Belokurov2018,Helmi2018GaiaEnceladus}, and 
Sequoia \citep{Myeong2019}.

Clearly, the GCs that are currently observed to belong to the MW are important probes of the formation and evolution of the early Galaxy; in particular, the formation history of the outer halo is imprinted in these GCs, which are located at large Galactocentric distances. From a practical perspective, one of the best ways to investigate the origin of any given stellar system is to  
perform ``chemical tagging", a method that we employ here.  

Our target, Palomar~13 (hereafter Pal~13), is a stellar system located in the outer halo, at a Galactocentric distance of 27 kpc \citep{Siegel2001}. 
The cluster is is currently thought to be close to apogalacticon \citep{Kuepper2011}. 
By analyzing ages and dynamics of the Galactic GC system, \citet{Massari2019} associated Pal~13 with the 
Sequoia event, a metal-poor system ([Fe/H] peaking at $-$1.6 dex)  that merged with the MW 9 Gyr ago.
\citet{Cote2002} carried out a dynamical study of the cluster based on radial velocities measured from spectra with high resolution, but low signal-to-noise ratio 
(S/N) of candidate red giant branch 
(RGB) stars. 
They found an intrinsic velocity dispersion of 2.2$\pm$0.4 km~s$^{-1}$ , which, though modest, translates into a surprisingly high mass-to-light 
ratio. It is
possible, however, that this value may be inflated by spectroscopic binaries \citep{Blecha2004}. Alternatively, an inflated velocity dispersion 
could be 
the result of tidal heating or shocking \citep{Kuepper2011,Yepez2019}. A break in the cluster density profile has indeed been noted by some 
researchers, which manifests itself in a 
shallow outer surface density profile \citep{Cote2002,Bradford2011,Hamren2013,Munoz2018}, 

Here, we continue our efforts to obtain chemical abundance constraints of remote systems in the outer Galactic halo by 
performing a chemical analysis by coadding spectra of high spectral resolution but low S/N. 
This method was shown to be successful in our series of papers dealing with the group of Palomar clusters 
(\citealt{Koch2009,Koch2010,Koch2017Pal5}; see also \citealt{Koch2019Pal15}). 

This paper is organized as follows. In Sect. 2 we introduce the spectroscopic sample and the observations, while Sect. 3 
focuses on the radial velocity measurements and cluster membership. In Sect. 4 we describe the abundance analysis through spectrum coaddition. 
The respective results are presented in Sect. 5 before we discuss Pal~13 in a broader accretion context in Sect. 6. Finally, in Sect. 7 we summarize our findings.
\section{Targets and observations}
The data used in our work are part of the program described by \citet{Cote2002}, which studied the internal dynamics of outer halo GCs.
Spectra of 30 RGB candidates were taken between August 1998 and August 1999 with the High Resolution Echelle Spectrometer (HIRES) 
on the Keck I telescope \citep{Vogt1994} with the C1 decker (0.86$\arcsec$) and 1$\times$2 binning. The resulting spectral resolution is 
45000, and our spectra cover a wavelength range of 4300--6720\AA.

Target stars were selected by \citet{Cote2002} from the published color-magnitude diagrams (CMDs) of \citet[][ herafter ORS]{Ortolani1985}
and their own photometry, collected with Keck and the Canada-France-Hawaii Telescope (CFHT). Details for the stars we used are listed in Table~1,
and a CMD is shown in Fig.~1.
\begin{table*}[htb!]
\caption{Stellar properties}
\centering
\begin{tabular}{rcccccccccc}
\hline\hline       
 & $\alpha$  & $\delta$  & $t_{\rm exp}$  & S/N\tablefootmark{b} & B$-$V & V & v$_{\rm HC}$ & T$_{\rm eff}$  &  $\xi$  & \\
\rb{Star\tablefootmark{a}} & (J2000.0) &  (J2000.0) &  [s] & [px$^{-1}$]  & [mag] & [mag] & [km\,s$^{-1}$]  & [K]  &   [km\,s$^{-1}$] & \rb{log\,$g$}\\
\hline                  
ORS-118  & 23 06 50.09  & 12 47 13.79  & 1590 &           16  &   0.93 &  17.00  &   24.92$\pm$0.21      &   4749  &  1.98  &  1.75   \\
 ORS-72  & 23 06 48.51  & 12 46 19.02  & 2640 &           14  &   0.85 &  17.64  &   28.77$\pm$0.28      &   4935  &  1.80  &  2.10   \\
 ORS-31  & 23 06 49.96  & 12 45 27.69  & 1650 &           10  &   0.83 &  17.76  &   25.07$\pm$0.37      &   4988  &  1.75  &  2.18   \\
 ORS-91\tablefootmark{c} & 23 06 43.14  & 12 46 33.93  & 1650 &           10  &   0.67 &  17.81  &   24.60$\pm$0.64      &   5471  &  1.90  &  2.28   \\
 ORS-32  & 23 06 42.00  & 12 45 26.18  & 4200 &            7  &   0.89 &  18.02  &   19.67$\pm$0.27      &   4832  &  1.56  &  2.60   \\
    d41  & 23 06 48.29  & 12 45 46.21  & 2020 &            6  &   0.76 &  18.59  &   19.26$\pm$0.49      &   5186  &  1.59  &  2.67   \\
 ORS-86  & 23 06 44.70  & 12 46 29.56  & 1800 &            7  &   0.77 &  18.81  &   24.47$\pm$0.90      &   5157  &  1.53  &  2.77   \\
 ORS-36  & 23 06 43.84  & 12 45 35.75  & 1800 &            6  &   0.75 &  18.98  &   25.29$\pm$0.99      &   5216  &  1.56  &  2.77   \\
 ORS-63  & 23 06 44.45  & 12 46 10.01  & 3900 &            9  &   0.76 &  19.02  &   25.07$\pm$0.44      &   5186  &  1.44  &  2.83   \\
 ORS-87  & 23 06 44.82  & 12 46 30.05  & 1800 &            6  &   0.72 &  19.05  &   26.55$\pm$0.69      &   5309  &  1.47  &  2.91   \\
    910  & 23 06 39.38  & 12 47 35.35  & \phantom{1}600 &  3  &   0.73 &  19.28  &   23.77$\pm$0.69      &   5277  &  1.47  &  3.03   \\
    915  & 23 06 43.98  & 12 46 19.86  & 1800 &            5  &   0.73 &  19.58  &   24.69$\pm$0.69      &   5277  &  1.47  &  3.05   \\
 ORS-18  & 23 06 39.99  & 12 44 47.33  & 1000 &            4  &   0.73 &  19.62  &   25.16$\pm$0.69      &   5277  &  1.50  &  3.05   \\
 ORS-38  & 23 06 45.61  & 12 45 39.42  & 3400 &            4  &   0.74 &  19.65  &   21.72$\pm$0.78      &   5247  &  1.44  &  3.09   \\
    931  & 23 06 43.41  & 12 46 07.69  & 2100 &            5  &   0.72 &  19.70  &   25.45$\pm$0.77      &   5309  &  1.44  &  3.10   \\
 ORS-78  & 23 06 45.69  & 12 46 21.48  & 1000 &            3  &   0.72 &  19.73  &   22.34$\pm$1.71      &   5309  &  1.44  &  3.11   \\
 ORS-50  & 23 06 45.01  & 12 45 55.95  & 2700 &            3  &   0.72 &  19.76  &   24.15$\pm$0.93      &   5309  &  1.44  &  3.15   \\
  ORS-5  & 23 06 45.37  & 12 44 21.03  & 2700 &            3  &   0.72 &  19.86  &   20.95$\pm$0.90      &   5309  &  1.29  &  2.40  \\
\hline                  
\end{tabular}
\tablefoot{
\tablefoottext{a}{Stellar IDs from \citet{Ortolani1985} (ORS) and \citet{Cote2002}.}
\tablefoottext{b}{Given at 6600~\AA.}
\tablefoottext{c}{AGB star.  
Neither foreground stars nor the variable star V2 from \citet{Cote2002} are listed here because they were not included in the analysis.}
}
\end{table*}
\begin{figure}[htb!]
\begin{center}
\includegraphics[angle=0,width=1\hsize]{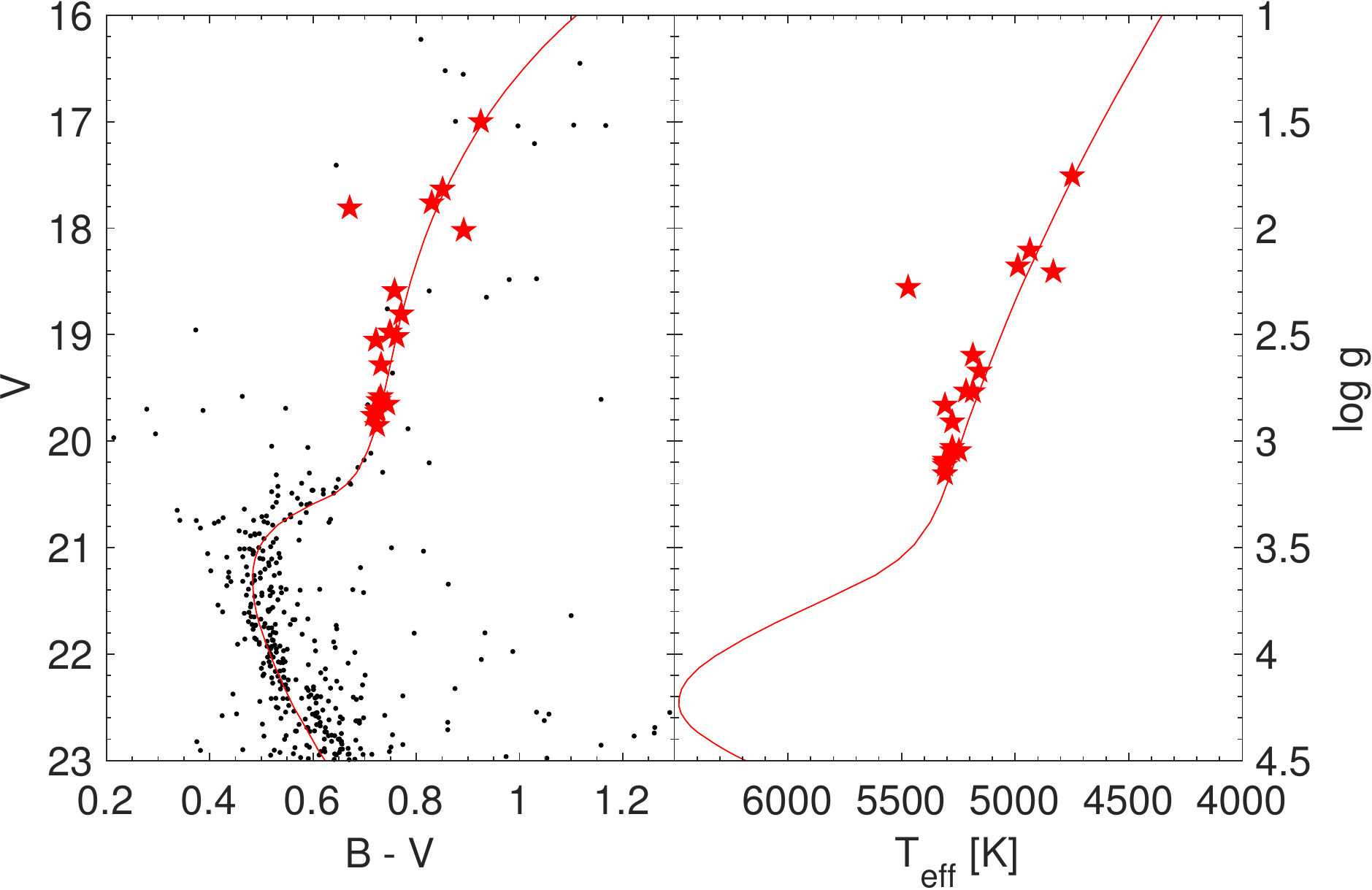}
\end{center}
\caption{CMD (left panel) and Kiel diagram (right panel). The member stars we used are shown as red symbols. 
Overlaid are metal-poor old isochrones from \citet{Bergbusch1992} using the GC parameters as shown in Fig.~4 of \citet{Cote2002}.}
\end{figure}

\section{Radial velocities and membership}
The radial velocity of each program star was measured by cross-correlating its spectrum against that of a master template created during each run from the observations of IAU standard stars. In order to minimize possible systematic effects, a master template for each observing run was derived from a similar and in some cases identical sample of IAU standard stars. From each cross-correlation function, we measured both $v_r$, the heliocentric radial velocity, and $R_{\rm TD}$, the \citet{TonryDavis1979} estimator of the
strength of the cross-correlation peak. We followed the procedures described in \citet{Vogt1995} to derive empirical estimates for radial velocity uncertainties using the measured $R_{\rm TD}$ values. The weighted radial velocities from \citet{Cote2002} are reported in Table 1. 
We note that additional radial velocity measurements for some of these stars have since been published by \citet{Blecha2004} and \citet{Bradford2011}.

For our abundance analysis, we reduced the HIRES data in a standard manner with the Makee\footnote{MAKEE was developed by T. A. 
Barlow specifically for reduction of Keck HIRES data. It is freely available on the World Wide Web at the Keck Observatory home page, https://
www2.keck.hawaii.edu/inst/common/makeewww} pipeline.
Specifically, we used the spectra of 18 radial velocity member stars with individual S/Ns of between 3 and 16 per pixel. Star V2 was excluded 
from our sample due to the difficulties inherent in abundance determinations of variable RR Lyrae stars unless the ephemerides are precisely 
known \citep[e.g.,][]{For2011}.  We also note that the member star ORS-91 is likely an asymptotic giant branch (AGB) star, judging by its 
position blueward of the RGB locus. \citet{Cote2002} discussed whether the stars ORS-32 and ORS-118 are members of Pal~13, based on 
CMD position, radial velocity, metallicity, and proper motion. The authors concluded that both these stars are probably bona fide members of Pal 13, and we therefore include them in our spectroscopic analysis. 
\section{Abundance analysis}
The main goal in acquiring this spectroscopic data set was to study the internal dynamics of faint GCs belonging to the MW halo. 
Individual exposure times were accordingly short, ranging from 10--70 minutes, which leads to the low S/Ns 
listed in Table~1. While these S/N levels are sufficient for measuring accurate radial velocities, they are too low for meaningful chemical 
abundance 
measurements, except for the overall metallicity. 
Therefore, we relied on our technique of coadding all available spectra and performing a coadded abundance analysis, which we have shown in 
previous studies to yield reliable results \citep{Koch2009,Koch2010,Koch2017Pal5}. 
\subsection{Stellar parameters}
We started by assigning each star an effective temperature from its (B$-$V) colors, for which we used the calibrations of \citet{Alonso1999}. 
These adopt an initial metallicity estimate of $-$2 dex, as found in previous CMD studies and a reddening of E(B$-$V)=0.10 \citep{Schlafly2011}. 
The typical uncertainty due to photometric and calibration errors on T$_{\rm eff}$ is $\sim$100~K.

The surface gravities were subsequently computed using the standard stellar structure equations 
(e.g., Eq.~1 in \citealt{Koch2008}). 
These use the above temperatures, a distance to Pal~13 of 26.9 kpc, an initial metallicity estimate of $-2$ dex 
 (which enters through the bolometric correction that we adopted from the Kurucz grids), 
and masses of 0.8 and 0.6 M$_{\odot}$ for RGB and AGB stars, respectively. We note, however, that the mass choice for the AGB star has no effect on the
final abundance results because its contribution to the coadded spectrum (i.e., one of 18 stars) is negligible. 
The errors on all input values are propagated through the formalism, 
which yields a typical uncertainty on log\,$g$ of 0.2 dex.

Microturbulence was fixed from the same empirical calibration of $\xi$ with T$_{\rm eff}$ that we obtained in our previous studies 
from the metal-poor halo sample of \citet{Roederer2014}. This relation reads $4.567 - 6.2694\,\times\,10^{-4}\, \times {\rm  T_{eff}}$. 
The  uncertainty on microturbulence is determined from the scatter about this above relation and amounts to 0.10 km\,s$^{-1}$.
 All resulting stellar parameters for our stellar atmospheres are listed in Table~1 and are summarized in the Kiel diagram of Figure~1 and in Figure~2. 
\begin{figure}[!tb]
\begin{center}
\includegraphics[angle=0,width=1\hsize]{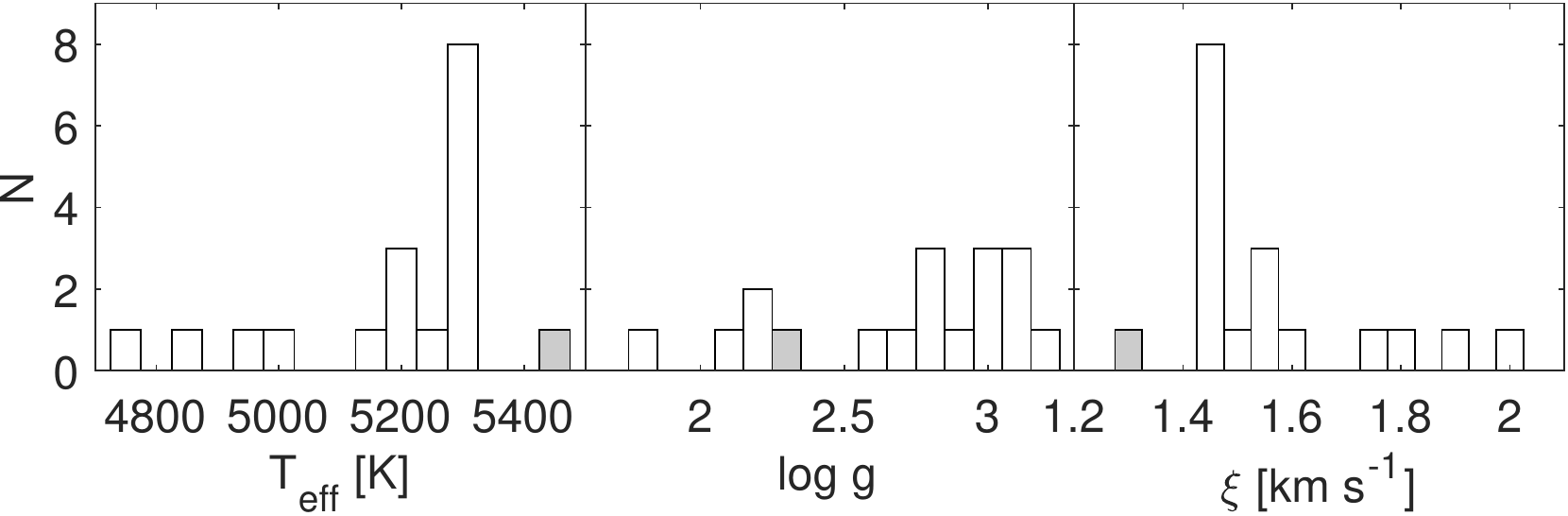}
\end{center}
\caption{Distribution of stellar parameters for the 18 target stars. Clear bars are for the RGB stars, and the solid bar indicates the AGB star.}
\end{figure}
\subsection{Coadded abundance measurements}
As in our previous studies, we began by median-combining the spectra of the 18 stars after weighting them by their S/Ns, 
which facilitates the $\sigma$ clipping within the IRAF program {\em scombine}. 
This results in an enhanced S/N of 33 per pixel at 6600~\AA, which is sufficient for a detailed element abundance study.  
On this composite spectrum, equivalent widths (EWs) were measured using the line lists from our previous studies, which in turn are based on 
\citet{Koch2016} and \citet{Ruchti2016}. In practice, we employed the IRAF task {\em splot} to fit Gaussian profiles to the individual lines.
These line lists and EW measurements are reported in Table~2.
\begin{table}[htb]
\caption{Line list}
\centering
\begin{tabular}{ccccc}
\hline\hline       
$\lambda$ &  & E.P. &  &  $\left<EW\right>$ \\
$[$\AA$]$ & \rb{Species} & [eV] &\rb{log\,$gf$} & [m\AA] \\
\hline                  
5688.205 &  Na\,{\sc i} & 2.10 &   $-$0.404  &  \phantom{3}23 \\
4571.096 &  Mg\,{\sc i} & 0.00 &   $-$5.623   & \phantom{3}78 \\
5172.700 &  Mg\,{\sc i} & 2.71 &  $-$0.402    & 315 \\
5183.600 &  Mg\,{\sc i} & 2.72 &   $-$0.180   &  365 \\
6155.130 &  Si\,{\sc i} & 5.61  &  $-$0.400   &  \phantom{3}31  \\
\hline                  
\end{tabular}
\tablefoot{Table~2 is available in its entirety in electronic form via the CDS.}
\end{table}

Throughout our analysis, we used the stellar abundance code MOOG \citep{Sneden1973}. Stellar atmospheres for every star were computed from the 
ATLAS grid of Kurucz one-dimensional 72-layer, plane-parallel, line-blanketed models without convective overshoot. 
In this analysis, we assumed that local thermodynamic equilibrium (LTE) holds for all species, and we used the $\alpha$-enhanced opacity distribution functions AODFNEW, 
which is expected to hold for metal-poor halo objects. 

Next, we computed theoretical EWs for all all measured transitions using the {\em ewfind} option of MOOG. These were combined into a mean $<$EW$>$ following the identical weighting 
scheme as for the observed spectra (Eq. 1 in \citealt{Koch2010}). Finally, the abundance ratios of the elements in question were varied so as to match the coadded $<$EW$>$
with the observed ones to deliver the integrated element abundance.    
\subsection{Abundance errors}
For our error analysis, we employed a standard procedure. 
To this end, we state the statistical error in terms of its 1$\sigma$ line-to-line scatter and N, the number of lines used to measure
every element abundance. 
For Na, Si, and Sc, only one line could be measured in each case. For Zn, we were able to measure two features that yielded a very low 1$\sigma$ scatter of lower than 0.01 dex, however.
In these cases, we varied the measured EWs by $\pm$5 m\AA, corresponding to $\sim$20\%, as a realistic 
uncertainty in our {\em splot} procedure. This yielded typical uncertainties of between 0.11 and 0.18 
dex for the Na, Si, Sc, and Zn abundance ratios, respectively.

In addition, we varied each stellar parameter by its representative uncertainty (T$_{\rm eff}\pm200$ K; $\log\,g\pm0.2$ dex; $\xi\pm0.1$ km\,s$^{-1}$)  
and redetermined abundances through our coaddition scheme. We also accounted for an uncertainty in the $\alpha$ enhancement when we switched 
to the solar-scaled opacity distribution functions ODFNEW. 
This procedure yielded the systematic errors reported in Table~3
in terms of the deviation of the abundances from the altered atmospheres from those based on the original parameter set.
\begin{table}[htb]
\caption{Systematic uncertainties. We list the deviations from the unaltered atmospheres. The last column shows the 
total  systematic error through addition in quadrature.}
\centering
\begin{tabular}{cccccc}
\hline\hline       
 & $\Delta$T$_{\rm eff}$ & $\Delta$log\,$g$  & $\Delta\xi$ &  &    \\
\raisebox{1.5ex}[-1.5ex]{Species} &
$\pm$$200$ K & $\pm$$0.2$ dex &  $\pm$$0.1$ km\,s$^{-1}$ & \raisebox{1.5ex}[-1.5ex]{ODF} & \raisebox{1.5ex}[-1.5ex]{$\sigma_{\rm sys}$}  \\
\hline                   
Fe\,{\sc i}  & $\pm$0.22 & $\mp$0.01 & $\pm$0.02 & \phs0.02 &  0.22 \\ 
Fe\,{\sc ii} & $\pm$0.07 & $\pm$0.09 & $\mp$0.01 &  $-$0.04 &  0.12  \\ 
Na\,{\sc i}  & $\pm$0.12 & $\mp$0.01 &   $<$0.01 & \phs0.01 &  0.12  \\ 
Mg\,{\sc i}  & $\pm$0.29 & $\mp$0.07 & $\mp$0.01 &  $-$0.01 &  0.30  \\ 
Si\,{\sc i}  & $\pm$0.06 & $\pm$0.01 & $\mp$0.01 &  $<$0.01 &  0.06  \\ 
Ca\,{\sc i}  & $\pm$0.16 & $\mp$0.01 & $\mp$0.02 & \phs0.01 &  0.17  \\ 
Sc\,{\sc ii} & $\pm$0.05 & $\pm$0.07 &     -0.01 &  $-$0.03 &  0.09  \\ 
Ti\,{\sc i}  & $\pm$0.27 & $\mp$0.01 & $\mp$0.02 & \phs0.02 &  0.28  \\ 
Ti\,{\sc ii} & $\pm$0.04 & $\pm$0.07 & $\mp$0.02 &  $-$0.04 &  0.10  \\ 
Cr\,{\sc i}  & $\pm$0.26 & $\mp$0.01 & $\mp$0.03 & \phs0.02 &  0.27  \\ 
Mn\,{\sc i}  & $\pm$0.24 & $\mp$0.01 &     -0.01 & \phs0.02 &  0.24  \\ 
Ni\,{\sc i}  & $\pm$0.19 &   $<$0.01 & $\mp$0.02 & \phs0.01 &  0.19  \\ 
Cu\,{\sc i}  & $\pm$0.25 &   $<$0.01 &   $<$0.01 & \phs0.01 &  0.25  \\ 
Zn\,{\sc i}  & $\pm$0.05 & $\pm$0.04 & $\mp$0.01 &  $-$0.01 &  0.06  \\ 
Y\,{\sc ii}  & $\pm$0.07 & $\pm$0.07 &     -0.01 &  $-$0.04 &  0.10  \\ 
Ba\,{\sc ii} & $\pm$0.10 & $\pm$0.07 & $\mp$0.04 &  $-$0.04 &  0.14  \\ 
\hline                  
\end{tabular}
\end{table}

As in past studies, we note that our approach assumes that all stars that enter the coaddition are subjected to 
identical errors. We also assumed that they share the same direction of deviations from the unchanged values. 
We further caution that the total systematic errors that we obtained by a simple quadratic sum of all contributions neglects
the covariances between the stellar parameters, predominantly between temperature and gravity, and, by our empirical scaling, the microturbulence. 
Thus, the values listed in Table~3 should be taken as conservative upper limits.

Additional error sources such as radial velocity shifts, potential foreground contaminants, and erroneous stellar type assignments were thoroughly discussed by 
\citet{Koch2010}. We reiterate that these effects account for less than $\sim$0.05 dex in the final error budget.
\section{Abundance results}
All abundance results and the errors as described in Sect. 4.3 are listed in Table~4.
These are placed into context with the MW halo, bulge, disks, and  MW GCs with a purported accretion origin 
 in Figs.~3 through 5. The sources for the GC abundance data are referenced in Table~5, and 
 their implications are further discussed in Sect.~6.
\begin{table}[htb]
\caption{Abundance results from coadded spectra. Abundance ratios for ionized species are given relative to Fe\,{\sc ii}. 
For iron itself, [Fe/H] is listed. The line-to-line scatter $\sigma$ and number of measured lines, N, determine the statistical 
error.}
\centering
\begin{tabular}{cccr|cccr}
\hline\hline       
Species & [X/Fe] & $\sigma$ & N & 
Species & [X/Fe] & $\sigma$ & N  \\
\hline                   
Fe\,{\sc i}  & $-$1.91 & 0.32   & 42 & Ti\,{\sc ii} & \phs0.28 & 0.20  &  6 \\  
Fe\,{\sc ii} & $-$1.90 & 0.30   &  7 & Cr\,{\sc i}  & $-$0.09 & 0.18   &  7 \\ 
Na\,{\sc i}  & \phs0.17 & \dots &  1 & Mn\,{\sc i}  & $-$0.06 & 0.16   &  7 \\ 
Mg\,{\sc i}  & \phs0.39 & 0.11  &  3 & Ni\,{\sc i}  &\phs0.08 & 0.26   &  6 \\ 
Si\,{\sc i}  & \phs0.43 & \dots &  1 & Cu\,{\sc i}  & $-$0.42 & 0.17   &  2 \\ 
Ca\,{\sc i}  &\phs 0.29 & 0.19  & 10 & Zn\,{\sc i}  & $-$0.20 & 0.01   &  2 \\ 
Sc\,{\sc ii} &  $-$0.12 & \dots &  1 & Y\,{\sc ii}  & $-$0.18 & 0.18   &  2 \\ 
Ti\,{\sc i}  & \phs0.24 & 0.14  &  7 & Ba\,{\sc ii} & $-$0.28 & 0.04   &  3 \\ 
\hline                  
\end{tabular}
\end{table}
\begin{table*}[htb]
\caption{List of GCs used in our comparison (Figs.~3--6). Only those GCs from the list of \citet{Massari2019} that have available abundance information are listed.}
\centering
\begin{tabular}{rrccr}
\hline\hline       
 & R$_{\rm GC}$ & [Fe/H] &   &  \\
\rb{ID} & [kpc] & [dex]  &\rb{Progenitor\tablefootmark{a}} & \rb{Abundance source} \\
\hline                  
NGC 2419 & 89.9 & $-$2.09 & Sgr & \citet{Cohen2012} \\  
NGC 5824 & 25.9 & $-$2.38 & Sgr & \citet{Roederer2016NGC5824} \\   
NGC 6715 & 18.9 & $-$1.51 & Sgr & \citet{Carretta2014} \\ 
Terzan 7 & 15.6 & $-$0.61 & Sgr & \citet{Sbordone2007} \\ 
Arp 2    & 21.4 & $-$1.80 & Sgr & \citet{Mottini2008} \\
Terzan 8 & 19.4 & $-$2.27 & Sgr & \citet{Carretta2014} \\   
Pal 12   & 15.8 & $-$0.82 & Sgr & \citet{Cohen2004} \\
%%%%%%%
\hline
NGC 5466 & 16.3 & $-$1.97 & Seq & \citet{Lamb2015} \\
 IC 4499 & 15.7 & $-$1.59 & Seq & \citet{DAlessandro2018} \\
NGC 7006 & 38.5 & $-$1.55 & Seq & \citet{Kraft1998} \\  
FSR 1758 & 11.8  & $-$1.58 & Seq & \citet{Villanova2019} \\
%%%%%%%
\hline    
 NGC \phantom{1}288 & 12.0 & $-$1.39 & G-E & \citet{Shetrone2000} \\
NGC  \phantom{1}362 & \phantom{1}9.4  & $-$1.17 & G-E & \citet{Carretta2013} \\   
NGC 1261 & 18.1 & $-$1.19 & G-E & \citet{Filler2012} \\   
NGC 1851 & 16.6 & $-$1.17     & G-E & \citet{Carretta2011} \\
NGC 1904 & 18.8 & $-$1.46     & G-E & \citet{Francois1991} \\
NGC 2298 & 15.8 & $-$1.91     & G-E & \citet{McWilliam1992} \\
NGC 2808 & 11.1 & $-$1.27     & G-E & \citet{Marino2017} \\
NGC 4833 & \phantom{1}7.0  & $-$2.25 & G-E & \citet{Roederer2015} \\
NGC 5897 & \phantom{1}7.4  & $-$2.04 & G-E & \citet{Koch5897} \\   
NGC 6205 & \phantom{1}8.4  & $-$1.50 & G-E & \citet{Cohen2005M3} \\
NGC 6229 & 29.8 & $-$1.13 & G-E & \citet{Johnson2017} \\   
Pal~15   & 38.4 & $-$1.94 & G-E & \citet{Koch2019Pal15} \\
NGC 6341 & \phantom{1}9.6  & $-$2.70 & G-E & \citet{Roederer2011} \\
NGC 6779 & \phantom{1}9.2  & $-$1.90 & G-E & \citet{Khamidullina2014} \\
NGC 6864 & 14.7 & $-$1.16 & G-E & \citet{Kacharov2013} \\  
NGC 7089 & 10.4 & $-$1.44 & G-E & \citet{Yong2014} \\  
NGC 7099 & \phantom{1}7.1  & $-$2.28 & G-E & \citet{Lovisi2013} \\
NGC 7492 & 25.3 & $-$1.82 & G-E & \citet{Cohen2005} \\
%%%%%%%
\hline
NGC 5272 & 12.0 & $-$1.39 & H99 & \citet{Cohen2005M3} \\ 
\hline                  
\end{tabular}
\tablefoot{\tablefoottext{a}{Following the assignments of \citet{Massari2019} to the Sgr dSph (Sgr), the Sequoia accretion event (Seq), 
Gaia-Enceladus (G-E), and the Helmi streams (H99).}
}
\end{table*}
\subsection{Iron abundance and metallicity}
We find a mean iron abundance from the neutral lines of  $-1.91\pm0.05$ (stat.) $\pm$ 0.22 (sys.) dex 
and an excellent ionization balance of [Fe\,{\sc i}/{\sc ii}]=$-$0.01$\pm$0.12. 
The latter was not necessarily fulfilled in our previous coadded abundance analyses of remote GCs, 
and the sense of the departure of Fe\,{\sc i} from Fe\,{\sc ii} was also found to vary from case to case 
(e.g., \citealt{Koch2009} vs. \citealt{Koch2010}). This implies that reaching ionization equilibrium is not 
a generic outcome  of our method, but rather depends on the quality of the adopted gravities and input (stellar) 
parameters. 

Older literature values for the  metallicity of Pal 13 from Str\"omgren photometry and low-resolution spectra range from 
$-1.67$ to $-1.90$ \citep{Canterna1978,Friel1982,Zinn1982}. This was later refined by 
\citet{Cote2002}, who coadded spectra of one star from the present sample (ORM-118) and measured an [Fe/H] of $-1.98\pm$0.31 from 28 Fe lines, but at a lower S/N than our present study. 
All these estimates are in excellent agreement within the errors with our high-resolution spectroscopic abundance.
We note, however, that the recent CMD study of \citet{Hamren2013} and a low-resolution spectroscopic analysis of 16 member stars
by \citet{Bradford2011} indicate a more moderate metallicity of $-1.6\pm0.1$ dex, which is marginally consistent with our result
if we account for the full statistical and systematic errors.
\subsection{Light elements (Na)}
Sodium is the only light element for which an abundance ratio could be obtained, and this result is only based on one detected line. 
One of the key features of interest in GCs is the presence of light-element variations and (anti-) correlations 
\citep{Carretta2009NaO,Bastian2018} due to the presence of multiple stellar populations. Unfortunately, none of the other 
participating elements (O and Al) could be measured so that our data cannot probe the respective internal processes
in Pal~13. 
Furthermore, as shown in \citet{Koch2017Pal5}, a coadded abundance analysis is highly insensitive 
 to information of intrinsic abundance spreads or correlations \citep[cf.][]{Bastian2019}.  
 We therefore do not pursue the otherwise regular halo-like [Na/Fe] of 0.17 dex any further.
\subsection{$\alpha$-elements (Mg, Si, Ca, and Ti)}
Manganese and silicon follow the canonical trend seen in metal-poor field and GC stars in that they lie directly at
 the plateau value of $\sim$0.4 dex. In contrast,  Ca and Ti take on values that are slightly lower than the halo mean, at 
 [Ca,Ti/Fe]=0.29 and 0.24 dex, respectively. 
It is worth noting that a very good ionization balance is also reached for Ti, at [Ti\,{\sc i}/{\sc ii}]=$0.03\pm$0.10.
We investigate in Sect. 6 the mild depletions in these abundances in the context of 
the purported accretion origins of low-$\alpha$ GCs in more detail.
\begin{figure}[htb]
\begin{center}
\includegraphics[angle=0,width=1\hsize]{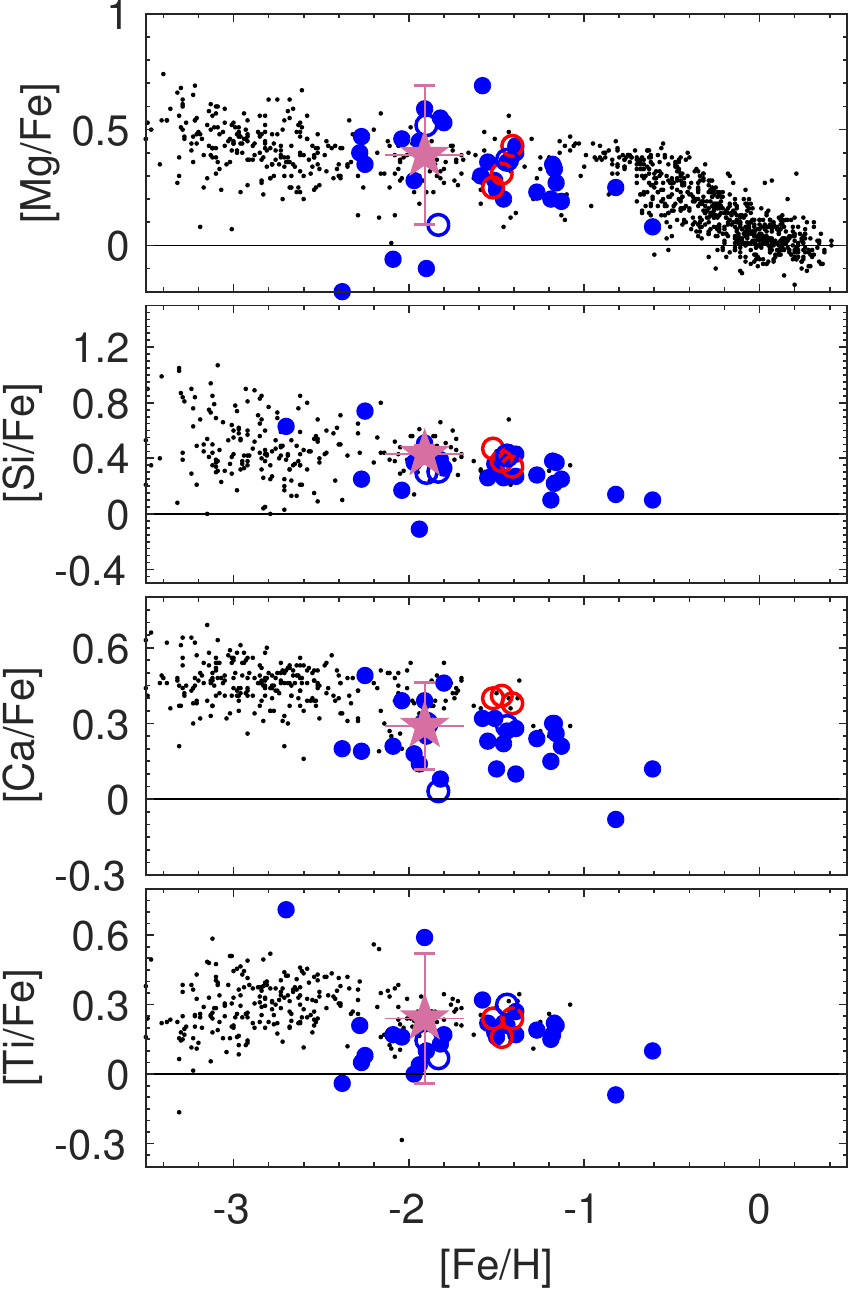}
\end{center}
\caption{Abundance results for the $\alpha$-elements. Literature data for the MW (black dots) are as follows: for the  
halo, \citet{Roederer2014}; for the bulge, \citet{Johnson2012,Johnson2014}; and for the disks, \citet{Bensby2014}.
Pal~13 is shown as the orchid star, where the error bar accounts for both statistical and systematic uncertainties. 
Blue filled points indicate the abundances of the comparison GCs (see Table~5 for references), 
while open blue symbols denote outer halo GCs beyond 20 kpc that have not been uniquely assigned to a 
specific progenitor. 
We also show as red circles the results from our systematic coadded abundance study of outer halo GCs \citep{Koch2009,Koch2010,Koch2017Pal5}.
}
\end{figure}

Owing to their production during different phases of the SNe~II \citep[e.g.,][]{Kobayashi2006}, 
such as the hydrostatic burning  in the progenitor (O, Mg), the explosion itself (Ca, Si), or 
an $\alpha$-rich freeze-out (Ti), the individual $\alpha$-elements
are not expected to uniquely trace one another.  
Here, we note that the  various ratio combinations  of [$\alpha$1/$\alpha$2] show values 
of about $-$0.15 to 0.10 dex in Pal~13, which are fully representative across the halo GCs that we use for reference purposes
in Sect. 6 (see also Figure~5 in \citealt{Koch2011}).

For simplicity, and bearing in mind the caveats of different origins, we chose to combine all 
four $\alpha$-elements into a single abundance ratio.
In doing so, we find a mean value of [$\alpha$/Fe] = 0.34$\pm$0.06 dex for for Pal~13.
\subsection{Fe-peak elements (Sc, Cr, Mn, Ni, and Zn)}
Our results for the Fe-peak elements are shown in Figure~4. 
We note that none of our own results was corrected for departures from LTE given the range in stellar parameters
in the coadded analysis method. During our spectral coaddition, differing non-LTE (NLTE) corrections would be smeared out (but at a level 
that is considerably lower than our quoted uncertainties). Furthermore, the largest corrections would arise chiefly for Cr
\citep{Bergemann2010Cr}, while the remainder of the elements differ only slightly \citep{Bergemann2008,Korotin2018}.
\begin{figure}[htb]
\begin{center}
\includegraphics[angle=0,width=1\hsize]{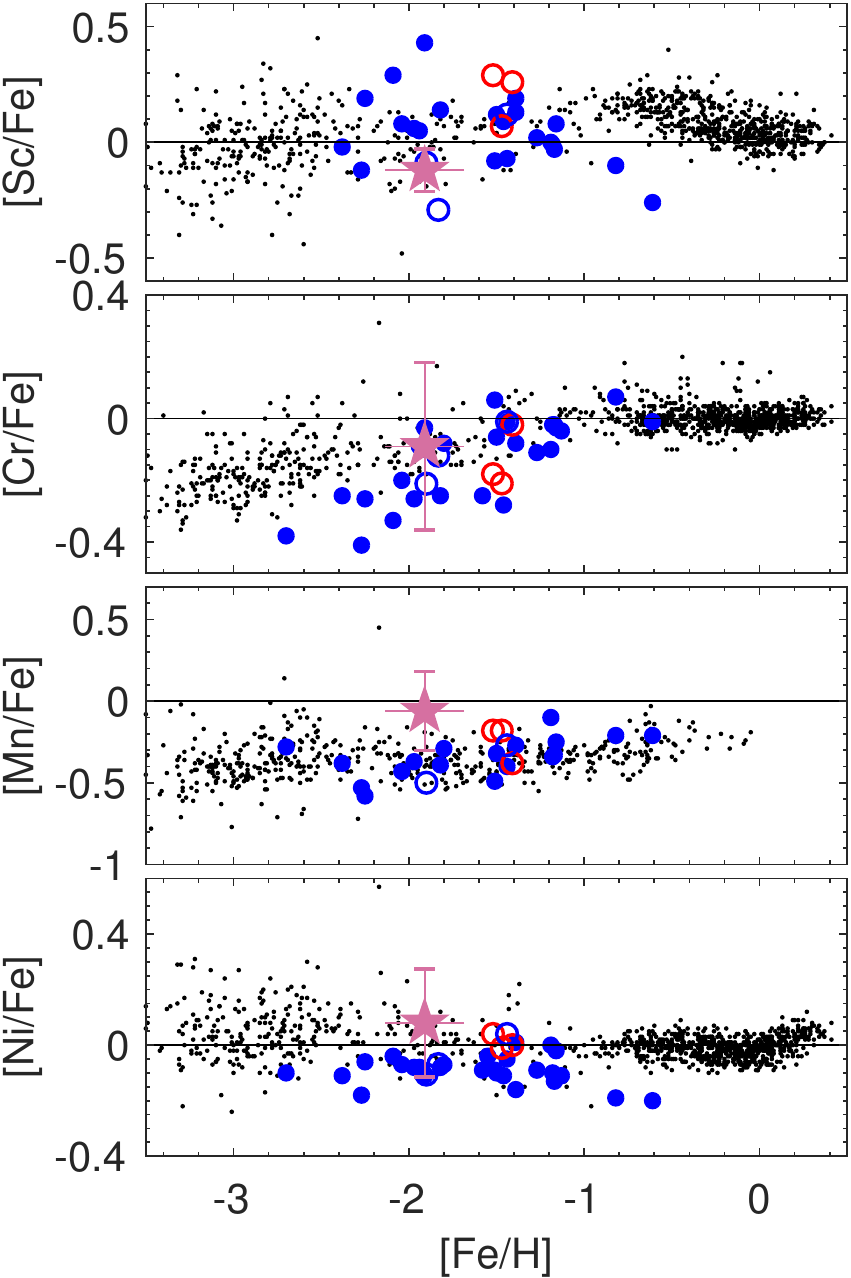}
\end{center}
\caption{Same as Figure~3, but for Fe-peak elements. Here, Sc and Mn abundances for the disk 
are from \citet{Battistini2015} and
\citet{Sobeck2006}, while Zn abundances for the bulge are taken from 
 \citet{Barbuy2015}.}
\end{figure}
None of these results holds any surprises, and the abundances of Pal~13 are consistent with those in halo stars
and GCs at similar metallicity. The majority of abundance ratios are consistent with solar or marginally subsolar
values. Mn is slightly enhanced relative to  typical halo and GC stars, but considering 
the uncertainty of 0.25 dex, 
we do not see an indication for any unusual Mn production in Pal~13. 
While the [Cu/Fe] ratio is depleted with respect to the solar value, this is still compatible 
with the low[Cu/Fe] branch occupied by halo stars. 
\subsection{$n$-capture elements (Y and Ba)}
\begin{figure}[htb]
\begin{center}
\includegraphics[angle=0,width=1\hsize]{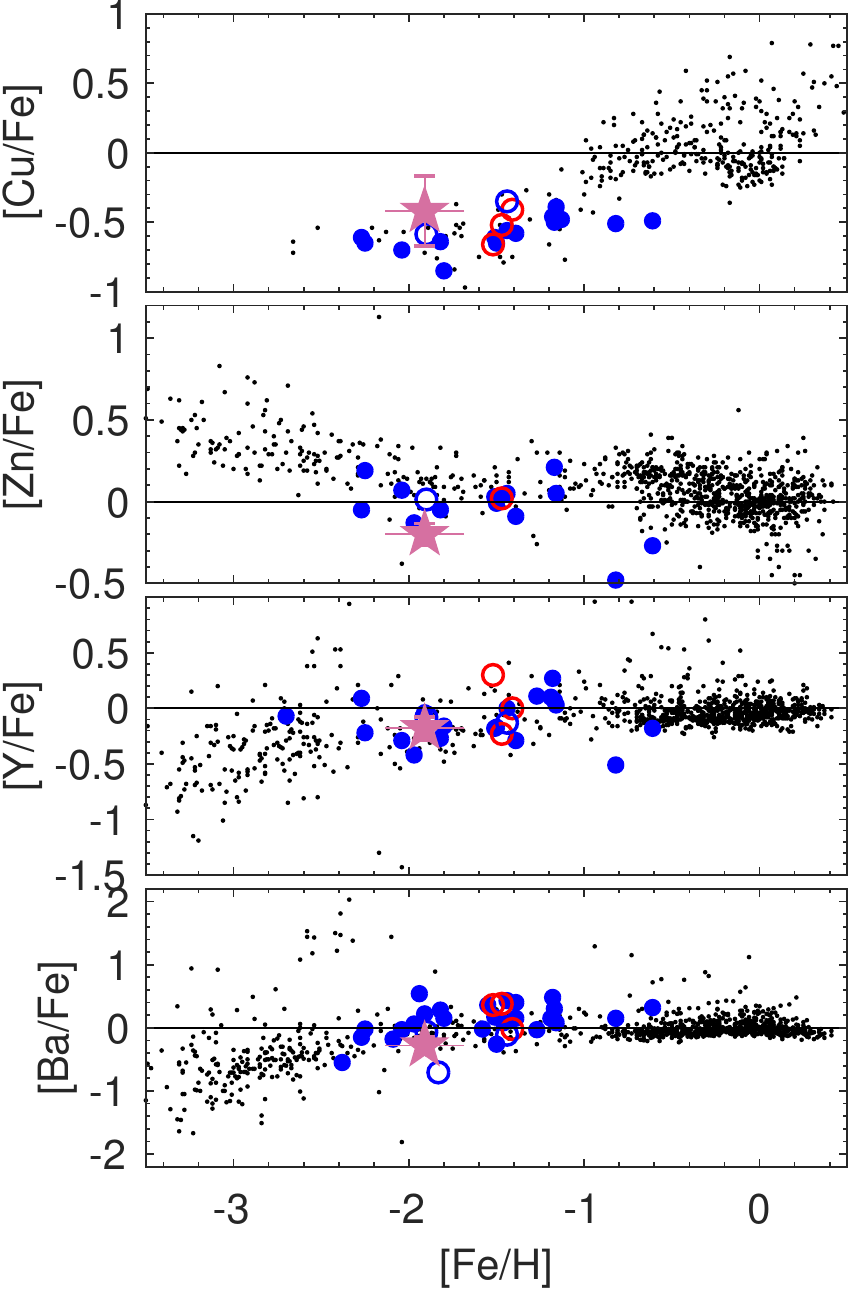}
\end{center}
\caption{Same as Figure~3, but for heavy and neutron-capture elements. 
Data for Cu across the MW are from \citet{Mishenina2002,Mishenina2011} and \citet{Johnson2014}.}
\end{figure}
The neutron-capture elements Y and Ba in Pal~13 are slightly subsolar (Fig.~5) and closely trace the 
underlying Galactic halo component. We note, however, that the Y abundance is based on only two detected lines that 
show a sizeable 1$\sigma$ dispersion. 
In the very metal-poor regime, the main source of the heavy elements is the $r$-process \citep{QianWasserburg2007}, 
although $s$-process signatures are seen as 
early as [Fe/H]=$-$2.6 \citep{Simmerer2004}. 
In particular, stars in dSph galaxies below $-$1.8 dex show systematically low [Y/Fe] abundances and consequently 
high [Ba/Y] ratios \citep{Venn2004,Tolstoy2009}. This was interpreted as an indication that the $r$-process production for Y in dSphs operates 
at a different site from that for Ba \citep{Pritzl2005}. 
At [Ba/Y] = $-$0.1 dex, the 
[$h$/$l$] ratio \citep[e.g.,][]{Cristallo2011} in Pal~13 is mildly depleted, but it can still be considered typical of metal-poor halo and GC stars.
\citet{Pritzl2005} noted  that in comparison with thick disk objects as well,
halo GCs tend to lie toward higher [Ba/Y] ratio, similar to stars in dSphs.  
While this holds true for a handful of halo objects at metallicities around $-$2 dex, Pal~13 does not exhibit such an offset. 
\section{Is Pal~13 an accreted object?}
The abundance of Pal~13 appears to be inconspicuous and very similar to the halo field and other
metal-poor GCs. Often, lower $\alpha$-abundances are taken as a sign of an accretion origin, although 
the presence of light-element (such as Mg) variations can partly veil this picture \citep[e.g.,][]{Pritzl2005,Carretta2017}.
In Pal~13, a mild depletion is seen in Ca and Ti, and from this point of view, we cannot unambiguously exclude an accretion origin for these elements. 
However, the overall [$\alpha$/Fe]  and the halo-like, enhanced Mg render any such evidence only  marginal.
How can we then establish from the abundances whether Pal~13 is an accreted object and even conclude on the type of progenitor?
This reasoning precludes that the GCs  follow the mean abundance of the underlying (Galactic or extra-Galactic) 
component, which is indeed observed \citep[e.g.,][]{Hendricks2016}. 
The situation is clearer for GCs that are associated with more complex systems such as the massive Sgr dSph stream system 
with its very broad metallicity distribution and spatial gradients \citep{Chou2007,Monaco2007,Hyde2015,CJHansen2018}.
\subsection{GC comparison sample}
The MW has evidently experienced at least three major accretion events that donated a major part of the stellar halo and also transferred their GC systems. 
The three the most recently identified accretion events (through ages and/or chemodynamics) are the disrupting Sgr dSph, the Sequoia event, and Gaia-Enceladus\footnote{The latter two 
have by now been identified with previously detected Galactic features, the nature of which 
had been  disputed in the past. In this vein, Gaia-Enceladus (also known as the Gaia sausage; \citealt{Belokurov2018}) 
can be identified with the Canis Major feature of the outer disk \citep{Momany2006}, whereas
Sequoia poses the observational evidence for the Kraken event postulated by \citet{Kruijssen2019}.}.
 
 \citet{Massari2019} combined the kinematics and ages for 151 GCs and asserted that 
40\% were formed in situ, 19\% stem from Gaia-Enceladus, and  5--6\% each originate in Sgr, Sequoia, and 
the Helmi streams \citep{Helmi1999,Koppelman2019}. 
\citet{Massari2019} further separated subsamples of GCs into ``main disk/bulge'' , that is, objects that are not associated with accretion events, but rather  
with the standard underlying MW components. Furthermore, these authors identified objects with high- and low-energy orbits that 
have no apparent accretion origin (``unassociated''). 
These ``regular''and presumably in situ GCs are ignored in our following reasoning. 
Furthermore, some of the MW GCs could not be unambiguously assigned to a single accretion event but are instead based on ages and kinematics. They 
share similar properties with several progenitors.
Thus, for our following comparison,  we are left with 18 GCs that are unequivocally associated with the Gaia-Enceladus event, 
7 from Sgr, one from the Helmi streams, and four from Sequoia, for which abundances are available in the literature.
Finally, we added 3 GCs that lie beyond 20 kpc from the literature. Detailed chemistry has been measured for these GCs, but
according to \citet{Massari2019}, they are not part of any of the substructures discussed above. These are 
NGC~5634 (R$_{\rm GC}$ = 21 kpc; \citealt{Carretta2017}), 
NGC~5694 (R$_{\rm GC}$ = 29 kpc; \citealt{Mucciarelli2013}), and 
Pal~14 (R$_{\rm GC}$ = 72 kpc; \citealt{Caliskan2012}). 
\subsection{Chemical tagging of Pal~13 and accreted GCs}
In Fig.~6 we plot the [$\alpha$/Fe] (bottom panels) and [Mg/Fe] (top panels) for the  
GCs for which abundance information is available. These ratios are representative tracers of a potential accretion origin.
We differentiate these into the three progenitor groups described above (Sect. 6.1) following the association scheme of 
 \citet{Massari2019}, and those at large distances without a unique progenitor.
Details and references for the chemical abundances we used in this comparison are given in Table~5. 
\begin{figure*}[htb]
\begin{center}
\includegraphics[angle=0,width=1\hsize]{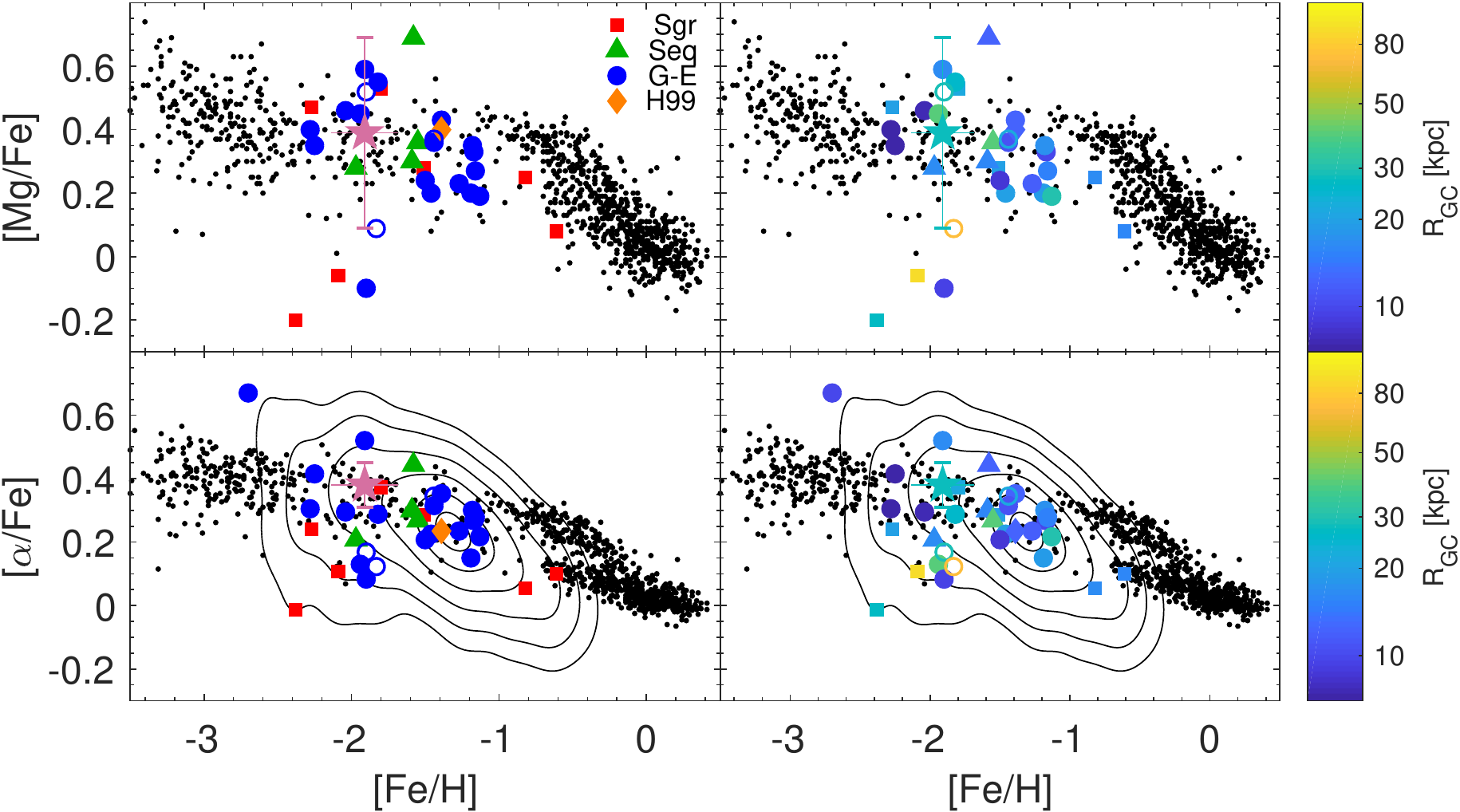}
\end{center}
\caption{Mg and $\alpha$-element ratios for Galactic stars, Pal~13 (star), and 
the GCs listed in Table~5, based on  \citet{Massari2019}. In the left panel, the clusters are 
 color-coded by their purported origin.
The right panels show the same data, but color-coded by (logarithmic) Galactocentric distance, maintaining the same symbols for the different progenitors.
Contours in the [$\alpha$/Fe] diagrams show the Gaia-Enceladus stellar component from \citet{Helmi2018GaiaEnceladus}
in terms of number densities from 0.5 to 3$\sigma$ in steps of 0.5$\sigma$. 
}
\end{figure*}

Other tracers of chemical evolution should, in the long run, also be employed to differentiate various progenitor groups 
and to characterize the GC-donating systems. One fo these diagnostics is the [Ba/Y] ratio, which is known to differ between  halo and dSph stars 
(\citealt{Pritzl2005}, see also Sect. 5.5.). However, heavy-element abundances are not known for a great number of reference GCs, and only 
one out of the four Sequoia-GCs has published [Ba/Y] abundance ratios. The same holds for the stellar component of these merger events, and 
future surveys and data releases, for example, of GALAH, are imperative to fill in this gap \citep{Buder2018}.

There is a caveat worth pointing out in this exercise: taking the mean abundance ratios as representative for the entire GC may not 
be appropriate for systems that have large 
abundance variations, or bimodalities to the point of spreading over more than 1 dex, as is the case for Mg in NGC 2419. 
Likewise, several abundance groups are found in NGC~7089 \citep{Yong2014}. 
Moreover, it is also not evident that all $\alpha$-elements should
be grouped into a single, straight average given their origin in different stages of the supernovae (SN)~II \citep{Venn2004,Kobayashi2006}. 
\subsubsection{Sagittarius?}
As the left panels of Fig.~6 indicate, the GCs associated with Sgr are scattered broadly in their metallicity, from below $-$2 up to about
 $-$0.7 dex.
This rather reflects the complex metallicity distribution of the entire Sgr system: as the most massive dSph satellite still visible in the MW system today,
it exhibits a broad age and metallicity range, including a very metal-poor stellar component \citep{CJHansen2018}.
The association of the GC with Sgr is mainly based on ages and kinematics, but two to three metal-poor GCs stand out in that they have 
significantly lower [$\alpha$/Fe] ratios: NGC~5824 and Pal~12. 
Of the Sgr candidates, only Arp~2 shares the fairly regular $\alpha$-abundances at the metallicity of Pal 13. 
We note that both GCs are located at fairly similar Galactocentric distances of 21 and 27 kpc, respectively.

Interestingly, NGC 5634 (at the same metallicity as Pal~13,  [Fe/H]=$-$1.9) has been tagged by \citet{Massari2019} as either a Helmi stream or 
a Gaia-Enceladus-based object, while
\citet{Law2010} added it to their list of Sagittarius clusters.
\subsubsection{Gaia-Enceladus?}
Both the stellar content of Gaia-Enceladus (shown as contours in the bottom panels of Fig.~6) 
and its associated GCs scatter broadly in abundance space. In this regard, finding a single
object such as Pal~13 in the locus occupied by this event is expected. 
This event contains both GCs with very low and very high $\alpha$ ratios, and the full range in between is covered.
If GCs do uniquely follow the underlying field component, then they should show a large overall scatter, or
a well-defined knee in the [$\alpha$/Fe] . The existence of low-$\alpha$ stars at low metallicities indicates low star formation efficiencies in the host system, 
leading to a (mass-dependent) early downturn in [$\alpha$/Fe] caused by the onset of the Fe-producing SNe~Ia \citep{Matteucci1990,Tolstoy2009}. 
If low-$\alpha$ stars are found at both low and high metallicities, this then appears to indicate a strongly varying star-forming efficiency and gradients
across the progenitor's main body, and if they exist, tidal streams.
The location of the $\alpha$ knee is now recognized to be dependent not only on galaxy mass, but also on  
the location 
within the galaxy \citep{Hendricks2014}. 

When we compare the entire chemical abundance pattern of Pal~13 with the named reference GCs, the closest match (in a $\chi^2$ sense)
is found for  NGC 6205 (M13). At first glance, this seems surprising because the object is
 more metal rich by 0.4 dex. Similarly, when we confine the comparison to the $\alpha$ elements alone among the Gaia-Enceladus GCs, then NGC 5897 
 provides the closest match.
 These similarities are, however, not consistent with the age and kinematical assignment of Pal 13 \citep{Massari2019}
 and do not provide firm evidence of a connection between Pal~13 and Gaia-Enceladus. 
\subsubsection{Sequoia?}
Based on their age and dynamical arguments, \citet{Massari2019} placed Pal~13 into the group with an origin in Sequoia, 
an object that was accreted 9 Gyr ago \citep{Myeong2019}, that is, at a time when Pal~13 was already some 3 Gyr old. 
Chemically, the Sequoia GCs do stand out, but they instead occupy a narrow metallicity window around $-$2 to $-$1.55 dex;  their $\alpha$ enhancements reach from mild depletions around 0.2 to overenhancements of about 0.5 dex (or 0.7 in Mg). 

Within the errors, Pal~13 resembles the most metal-poor progeny of Sequoia, NGC 5466. Intriguingly, \citet{Lamb2015} stated that 
they chose this object for abundance analysis in order to probe element trends that could establish a link with the Sgr system.
However, at its low metallicity ([Fe/H]=$-$2), the authors did not detect any obvious differences from the Galactic halo abundance patterns. 
This emphasizes the difficulties in uniquely tracing the origin of a system such as Pal~13 in a given merger event based on chemistry alone
(see also \citealt{ReinaCampos2019}).

\begin{figure}[htb]
\begin{center}
\includegraphics[angle=0,width=1\hsize]{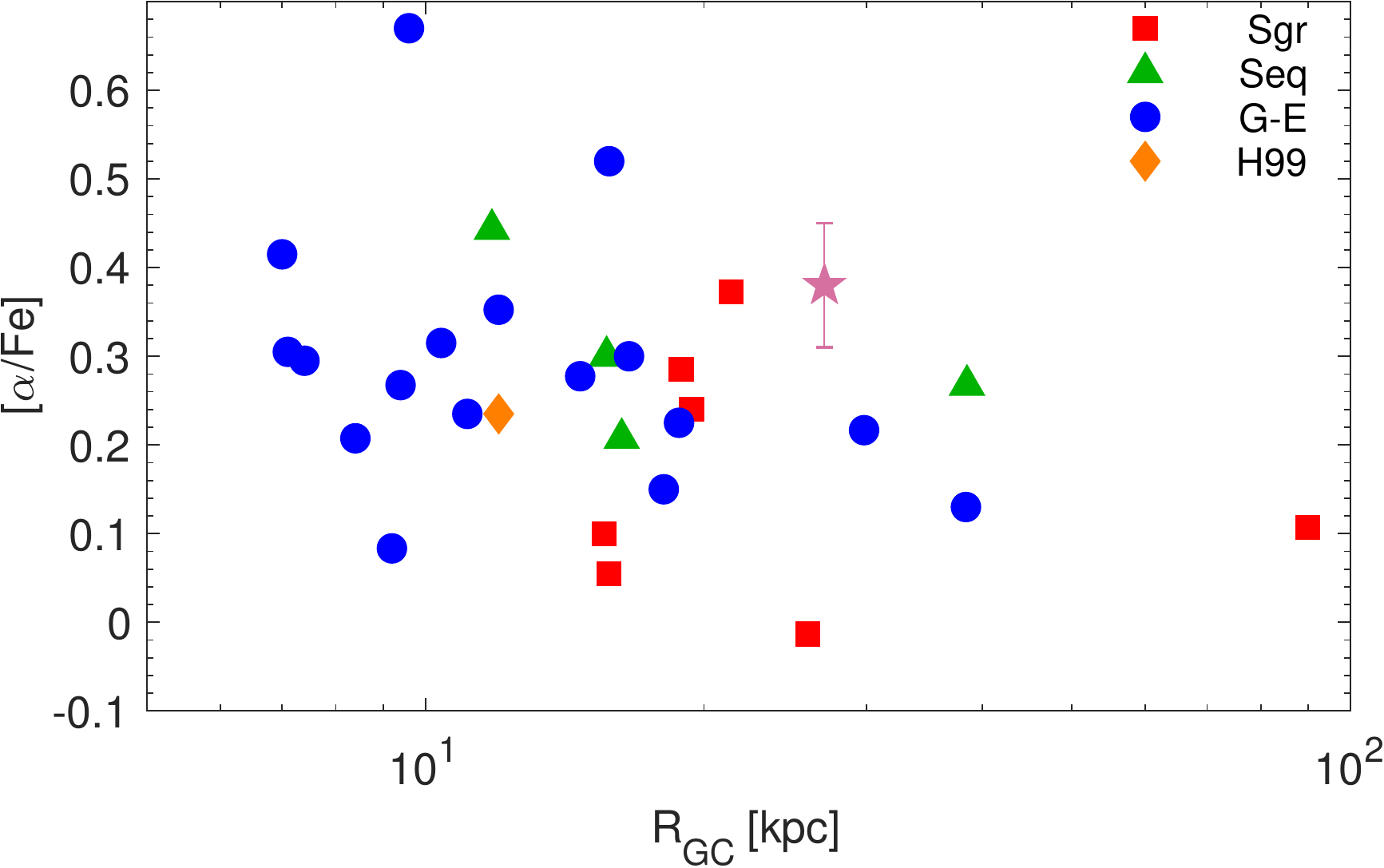}
\end{center}
\caption{Distance dependence of the [$\alpha$/Fe]  in the potentially accreted reference GCs.}
\end{figure}
Finally, we show in Fig.~7 the mean $\alpha$ enhancement of this sample of reference GCs as a function of Galactocentric distance
(see also the right panels of Fig.~6; cf. \citealt{Mackey2004}). Here, the ratios in the Sgr GCs cover a broad range; that is, there is essentially 
no dependence on distance, as shown by the Pearson $r$ value of $-0.18$. Although not
convincingly significant, it is still worth noticing that the correlation between   [$\alpha$/Fe] and R$_{\rm GC}$
is most pronounced for the Sequoia clusters (at $r=-0.40$), followed by the Gaia-Enceladus system with $r=-0.35$.
The latter, however, strongly depends on the two GCs with a strong $\alpha$ enhancement (NGC 2298 and 6341, the latter being the most metal-poor
GC in the MW halo). 

We also note that the majority of Sequoia GCs are currently at large Galactocentric distances in excess of 10 kpc, while those
from Gaia-Enceladus cover a range from $\sim7$ to 38 kpc,  40\% of which lie within 10 kpc. This may support 
the greater mass of this system, which caused it to perpetrate deeper into the young MW halo. 
\section{Summary and conclusions}
We have performed a coadded chemical abundance analysis of the outer halo GC Palomar~13 and found 
that the majority of its abundance patterns is highly regular. In the past, some unusual properties of this object (i.e., a higher than expected velocity dispersion and unusual surface density profile) have prompted suggestions that it has experienced tidal heating.
However, we found only few discernible peculiarities in chemical abundance space, although this might be expected given that its chemical patterns were likely established prior to any heating  (see also \citealt{Koch2017Pal5}).

Of higher importance is the question whether Pal 13 and other similar objects in the outer halo have an in  or  ex situ origin.
\citet{Kruijssen2019} and \citet{Massari2019} estimated that $\sim$40\% of the known MW GCs (i.e., $\sim$60 objects) have an extragalactic origin.
For some of these halo objects, the evidence is compelling that they have been accreted: that is, those that show clear chemical signatures reminiscent of 
dSph galaxies or dynamical pairings with known accretion events.
Other cases are less clear because they coincide with the main chemical abundance space occupied by Galactic halo stars. 
Chemically, Pal~13 belongs to the latter class. It has been suggested that it originated in the Sequoia event 9 Gyr ago, and both
its age and orbit support this conjecture. Its abundance ratios are very similar to those of NGC 5466, which has also been identified with 
Sequoia. On the other hand, the latter object is equally consistent with the MW halo distribution, and unambiguous evidence for
an accretion origin of either (outer) halo GC has not yet emerged.
%
%
%And so it was that, later, when the ... 
%
%
\begin{acknowledgements}
A.K. acknowledges financial support from the Sonderforschungsbereich SFB 881 
``The Milky Way System'' (subprojects A03, A05, A08) of the DFG. 
The authors are grateful to the anonymous referee for a swift report and to  J.M.D Kruijssen for helpful discussions.
This work was based on observations obtained at the W. M. Keck Observatory, which is operated jointly by the California Institute of Technology and the University of California. We are grateful to the W. M. Keck Foundation for their vision and generosity. We recognize the great importance of Mauna Kea to both the native Hawaiian and astronomical communities, and we are grateful for the opportunity to observe from this special place.
 \end{acknowledgements}
\bibliographystyle{aa} % style aa.bst
\bibliography{ms} % your references Yourfile.bib
\end{document}